\documentstyle[12pt]{article}
\makeatletter
\newcommand{\singlespacing}{\let\CS=\@currsize\renewcommand{\baselinestretch}
{1.0}\tiny\CS}
\newcommand{\doublespacing}{\let\CS=\@currsize\renewcommand{\baselinestretch}
{1.5}\tiny\CS}

\begin{document}

\begin{center}
{\large{Asymmetric Squares as Standing Waves in Rayleigh-B\'enard Convection}}\\
\vspace{1cm}
Alaka Das, Ujjal Ghosal and Krishna Kumar\\
\it{Physics and Applied Mathematics Unit, Indian Statistical Institute\\
203, Barrackpore Trunk Road, Calcutta-700 035, India}\\

\end{center}
\begin{abstract}

 Possibility of asymmetric square convection is investigated 
numerically using a few mode Lorenz-like model for thermal convection in 
Boussinesq fluids confined between stress free and conducting flat boundaries. 
For relatively large value of Rayleigh number, the stationary rolls become
unstable and asymmetric squares appear as standing waves at the onset
of secondary instability. Asymmetric squares, two dimensional rolls and
again asymmetric squares with their corners shifted by half a wavelength
form a stable limit cycle.\\

\end{abstract}

\vspace{1cm}

\ \ \  Two dimensional stationary roll-patterns are known to be the
only stable solutions at the onset of thermal convection in a thin layer
of Boussinesq fluid confined between conducting boundaries[1] except in the
case of fluids with vanishingly small Prandtl number[2]. The stationary
convection in the form symmetric square cells are unstable [1] except in
fluids of very high Prandtl number $\sigma$ at very high values of
Rayleigh number $R$ [3]. Assenheimer and Steinberg[4] observed
an interesting possibility of hexagonal convective cells with both up- and
down- flow in the center of the cell at $R$  approximately
twice the value of critical Rayleigh number $R_c$. The  possibility of dual
types of hexagonal convective cells was also established by Clever and Busse
[5]. Recently, Busse and Clever[6] predicted a new possibility of stationary
convection in the form of asymmetric square cells with both up- and down-
flow in the center for rigid and thermally conducting horizontal boundaries.

\vspace{1cm}
 In this work we construct and study a Lorenz-like model[7] that describes
 convection patterns in the form of rolls and squares, both symmetric as
 well as asymmetric ones, in a thin layer of Boussinesq fluid of moderate
 and high Prandtl numbers ($\sigma  > 2 $)
 at high values of $R$. The nonlinear superposition of mutually  
 perpendicular sets of rolls of the same wave number would be called  
 asymmetric squares when the structure (e.g., shadowgraph picture) does not 
 possess four-fold symmetry. It happens when the intensities of the two set 
 of rolls are different. The model allows us to study competetion between
 rolls and asymmetric squares. In the following we shall derive the model
 from hydrodynamic equations and investigate the model numerically for
 possible stable solutions. We then present the results and discuss them.

\ \ \   We consider an infinite layer of Boussinesq fluid
 of kinematic viscosity $\nu$, thermal diffusitivity $\kappa$ and 
 thickness $d$ confined between two perfectly conducting horizontal 
 boundaries and heated underneath. Using the length scale $d$, the time scale
 $\frac{d^2}{\kappa}$, and the temperature scale as the temperature difference
 $\Delta T$ between lower and upper boundaries, the non-dimensional form of
 hydrodynamic equations in Boussinesq approximation read
\begin{eqnarray}
\partial_t{\bf v}+({\bf v}.{\bf{\nabla}}){\bf v} &=& -{\bf{\nabla}}p +\sigma
(\theta{\bf{e_3}}+{\nabla}^2 {\bf v})\\
{\bf{\nabla}}.{\bf v} &=& 0\\
\partial_t\theta+{\bf v}.{\bf{\nabla}}\theta &=& {\nabla}^2\theta+R{\bf v}.{
\bf e_3}
\end{eqnarray}
 where ${\bf v}=(v_1,v_2,v_3)$ are the velocity fields, $\theta$ the deviation 
 from the conductive temperature profile and $p$ the deviation from static 
 pressure of conductive state due to convective instability. Prandtl number
 $\sigma$ and Rayleigh number R are defined as $\sigma = \frac{\nu}{\kappa}$
 and $R = \frac{\alpha(\Delta T)g{d}^3}{\nu\kappa}$, where $\alpha$ is the
 coefficient of thermal expansion of the fluid, g the acceleration due to
 gravity. The unit vector ${\bf e_3}$ is directed vertically upward, which
 is assumed to be the positive direction of $x_3$-axis. The boundary
 conditions at the stress-free conducting flat surfaces imply
 $\theta = v_3 = \partial_{33}{v_3} = 0$ at $x_3$=0, 1. Taking curl twice of
 the momentum equation (Eq. 1) and using the continuity condition (Eq. 2),
 the equation for vertical velocity reads \\
\begin{equation}
\partial_t{\nabla}^2v_3 =\sigma{\nabla}^4v_3 +\sigma{{\nabla}_H}^2\theta - 
{\bf{e_3}}.{\bf{\nabla}}\times[({\bf{\omega}}.{\bf {\nabla}}){\bf v} - 
({\bf v}.{\bf{\nabla}}){\bf{\omega}}] \\
\end{equation}
 where ${\bf{\omega}} = {\nabla}{\times}{\bf v}$  is the vorticity, and
 ${{\nabla}_H}^2 = \partial_{11} + \partial_{22}$ is the horizontal Laplacian.

\ \ \ We employ the standard Galerkin procedure to
 describe the convection patterns at the onset of secondary instability in
 relatively high Prandtl number fluids. The spatial dependence of all 
 vertical velocity and temperature field are expanded in a Fourier series,
 which is compatible with the stress-free flat conducting boundaries and
 periodic boundary conditions in the horizontal plane. Here, we restrict
 ourselves to standing patterns and, hence all time-dependent fourier
 amplitudes will be assumed to be real. The expansions for all the fields
 are truncated to describe straight cylindrical rolls and patterns arising
 from the nonlinear superposition of mutually perpendicular set of rolls of
 the same wavenumber. Perturbative fields with the same wavelength
 in mutually perpendicular directions are likely to occur in square 
 containers. The vertical velocity $v_3$ and $\theta$ then read as

  $$ v_3 = [W_{101}(t) cos(kx_1) + W_{011}(t) cos(kx_2)]sin(\pi x_3) + 
 W_{112}(t) cos(kx_1)cos(kx_2)sin(2{\pi}x_3) $$ 
 \begin{equation}
 + ...,\\
 \end{equation} 
 $$ \theta = [\Theta_{101}(t)cos(kx_1) + \Theta_{011}(t)cos(kx_2)]sin({\pi}x_3) 
 + \Theta_{112}(t)cos(kx_1)cos(kx_2)sin(2{\pi}x_3) $$ 
 \begin{equation}
 + \Theta_{002}(t) sin(2{\pi}{x_3}) + ....
 \end{equation}
 The horizontal components of the velocity field can easily be computed using
 the equation of continuity. Projecting Eq.(4) for $v_3$ and the equation for 
 $\theta$ (Eq. 3) on above modes, we arrive at the following minimal mode 
 Lorenz-like model
                
\begin{eqnarray}
\tau\dot{\bf{X}} &=& \sigma(-{\hat q}^2{\bf X}+\frac{{\hat k}^2}{{\hat q}^2}
{\bf Y})+\left(\begin{array}{c} X_{2}\\X_{1} \end{array} \right) S\\
\tau\dot{\bf{Y}} &=& -{\hat q}^2{\bf{Y}}+(r-Z) {\bf{X}} +\left(\begin{array}{c} 
X_{2}\\X_{1} \end{array}\right) T\\
\tau\dot{S} &=& -2\sigma{\hat d}^2 S + \sigma\frac{{\hat k}^2}{{\hat d}^2}
{T} - \frac{{\hat q}^2}{2{\hat d}^2} X_{1}X_{2}\\ 
\tau\dot{T} &=& -2{\hat d}^2{T}+r{S} - \frac{1}{4}(X_{1}Y_{2}+ X_{2}Y_{1})\\
\tau\dot{Z} &=& -b Z+{\bf{X}}.{\bf{Y}}
\end{eqnarray}
where the linear modes ${\bf X}= \left(\begin{array}{c} X_1\\X_2 \end{array}
\right) = \frac{\pi}{\sqrt 2 q_c^2} \left(\begin{array}{c} {W_{101}}\\{W_{011}}
\end{array} \right)$~and
${\bf Y}= \left(\begin{array}{c} Y_1\\Y_2 \end{array} \right) = \frac{{k_c}^2} 
{2{\sqrt 2} q_c^5} \left(\begin{array}{c} {\Theta_{101}}\\{\Theta_{011}} 
\end{array} \right)$ represent  the vertical velocity and the temperature
field respectively. The nonlinear mode $Z = \frac{-\pi k_c^2}{q_c^6}\Theta_
{002}$ denotes heat  flux across the fluid layer. Modes $S = \frac
{\pi}{4q_c^2}W_{112}$ and $T = \frac{\pi k_c^2}{4 q_c^6}\Theta_{112}$ are 
essential to describe nonlinear coupling between two sets of mutually 
perpendicular rolls. The parameters defined by   
$k^2 = \pi^2 + k^2,~~ {\hat q}^2 = \frac{q^2}{q_c^2},~~ {\hat k}^2 = \frac
{k^2}{k_c^2},~~{\hat d}^2 = \frac{2{\pi}^2 + k^2}{q_c^2}$ are, in general, 
wavenumber dependent, while  other parameters given by
$ k_c^2 = \frac{\pi ^2}{2},~~ q_c^2 = \pi^2 + k_c^2,~~ \tau = \frac
{1}{q_c^2},~~ b= \frac{4{\pi}^2}{q_c^2}$ are constants in the model. The 
reduced Rayleigh number $r = \frac{Rk_c^2}{q_c^6}$ is the control parameter of
the problem. We set, hereafter, $k = k_c$, the wavenumber of stationary rolls 
at the onset of primary instability. This makes $\hat k^2 = \hat q^2 = 1$, 
$\hat d^2 = \frac{5}{3}$ and $b= \frac{8}{3}$  in our model for convection.

\ \ \ The steady two dimensional(2D) rolls parallel to $x_1$-axis is given by $X_1 = 0 = Y_1$. This 
makes the nonlinear modes $S$ and $T$ to decouple from the system. Our model
then reduces to well known Lorenz model with steady solutions given by 
$X_2 = Y_2 = \sqrt{b(r-1)}$ and $Z = r-1$. However, these solutions become
unstable at much lower values of reduced Rayleigh number $r$ than that 
predicted by the original Lorenz model[7]. The lower curve in the Figure 1
shows the critical value $r_o$ of reduced Rayleigh number, at which the 2D
rolls become unstable, as a function of the Prandtl number $\sigma$. The
critical value $r_o$ increases with increasing $\sigma$. \\

\ \ \ The steady and perfect squares are obtained by setting  $X^2_1 = X^2_2 $ in
our model. Such solutions are given by,\\
$ Y_2^2=Y_1^2,~~ Y_1 = -\frac{2 \sigma (r-4{\hat d}^6) r X_1 + r X_1^3}
{2 \sigma {\hat d}^4X_1^2 -2 \sigma(r-4{\hat d}^6)(1+\frac{2X_1^2}{b})},~~
S=\frac{sign(X_1X_2)}{\sigma(r-4{\hat d}^6)}(\frac{\sigma}{4}{\bf X.Y} +
{\hat d}^2 X_1^2),$\\
$T= \frac{sign(X_1X_2)}{2 \sigma (r-4{\hat d}^6)}(\sigma {\hat d}^4 {\bf X.Y}
+ rX_1^2),~~ Z=\frac{1}{b}({\bf X.Y})$, 
where $X_1$ is a possible real root of the algebraic equation\\
$X_1^4 + AX_1^2 + B = 0$\\
with $A$ = $\frac{2 \sigma^2 {\hat d}^4 -\frac{4 \sigma^2}{b} + 2{\hat d}^2
+ 2 \sigma r}{(\frac{4{\hat d}^2}{b} + \frac{1}{2})}$ and $B=\frac{2 \sigma^2
(r - 4{\hat d}^6)(r-1)}{(\frac{4{\hat d}^2}{b} + \frac{1}{2})}$.
For $1<r<r_o(=4{\hat d}^6)$, which defines the validity range of the model,
$B$ is negative and consequently positive root of $X_1^2$ exists. This
implies that the convective patterns in the form of perfect squares exist.
However, they are found to be always unstable.

\ \ \ We study the time-dependent solutions by numerically integrating
our model. We investigate the dynamics of the patterns as a function of
increasing $r$ in two ways- first by using the data from previous solution
as initial conditions for a new r; and secondly by choosing randomly small
values of all variables as initial conditions. We get the same result with
suitably chosen initial conditions. For $r_o < r < r_{ir} $ (see Fig. 1),
we find oscillatory solutions. The stationary rolls become  unstable and a
new set of rolls perpendicular of the old one develops. The competition
between these two sets of rolls leads to  a time periodic sequence of
rolls and patterns arising from nonlinear superposition of two sets of rolls. 
For $r \ge r_{ir}$, the solutions become irregular in time indicatiing more
modes are required to study this regime of parameter space.

\ \ \ Figure 2 shows the temporal behaviour of the  
amplitude $X_1$ of the new set of rolls as well as the amplitude $X_2$ of the 
old set of rolls. While $X_1$ oscillates with zero mean, $X_2$ oscillates 
around a finite value. The time period of $X_1$ is always double of that of 
$X_2$. In fact all the modes $(X_1, Y_1, S, T)$ that develop at $r > r_o$ 
oscillate with same period and with zero mean. The periods of $Y_2$ and $Z$
are the same as that of $X_2$. The frequency $f=1/T$ of oscillation of $X_2$
at the onset  of oscillatory instability increases with increasing Prandtl
number (see Fig. 3).

\ \ \ The projection of limit cycle on $Y_1 - Y_2$ plane is shown in Fig. 4. The
mean of oscillating amplitude of 2D roll mode ($Y_2$) decreases and the 
perturbative amplitude ($Y_1$) increases with increasing $r$. Figure 5 shows
the sequence of shadowgraphs of the patterns for $r_o < r < r_{ir}$. The
intensity of old 2D rolls decreases when a new set of rolls perpendicular
to the old one appears. While the spatial positions for up and down flows
in the old set remain fixed, these  positions alternate for new set of rolls. Consequently,
the corner positions of square patterns slide back and forth along the set of 
old rolls.  The standing waves in form of square patterns do not have
four-fold symmetry although they preserve inversion symmetry. The time
periodic appearances of asymmetric squares, 2D rolls, asymmetric squares 
again with shifted positions of the maximum up or down flows by half a
wavelength and 2D rolls form a limit cycle. Although the competetion between
rolls and symmetric squares is known in binary mixtures[8, 9], but time
periodic sequence of asymmetric squares and two dimensional rolls in pure
fluids is qualitatively new.  
 
\ \ \ We now test the stability of this limit cycle by introducing the vertical 
vorticity in the model. We expand the vertical vorticity $\omega_3$ as 
$$ \omega_3 = [\zeta_{101}(t) cos(kx_1) + \zeta_{011}(t) cos(kx_2)] cos(\pi x_3) + 
\zeta_{110}(t) cos(kx_1)cos(kx_2) $$ 
\begin{equation}
+ ...,\\
\end{equation}
Simultaneously, for the consistency of the model, we add a mode
$W_{{\bar 1}{\bar 1} 2}$ in the expansion of vertical velocity $v_3$, and
a similar mode $\Theta_{{\bar 1}{\bar 1} 2}$ in the expansion of temperature
field $\theta$. This makes a Lorenz-like model consisting of twelve modes.
The results of this model for fluids with $\sigma > 2 $ exactly reproduces
the results of the seven mode model discussed earlier. Only for $\sigma < 2$
and much higher values or $r$, vertical vorticity is excited. So for
$\sigma >2 $ and perturbations of wavenumber same as that of 2D rolls, we
always see a stable limit cycle. The results of our model indicate that for
fluids with moderate and large values of Prandtl number enclosed preferably
in a square container, the time periodic competition between the asymmetric
squares and rolls  should be realizable.
 
Acknowledgement:We acknowledge support from DST, India. U. Ghosal, a summer
trainee from I I T, Kharagpur, acknowledges partial support from I S I,
Calcutta. 

\newpage
\begin{center}
{\bf FIGURES}
\end{center}

\noindent 
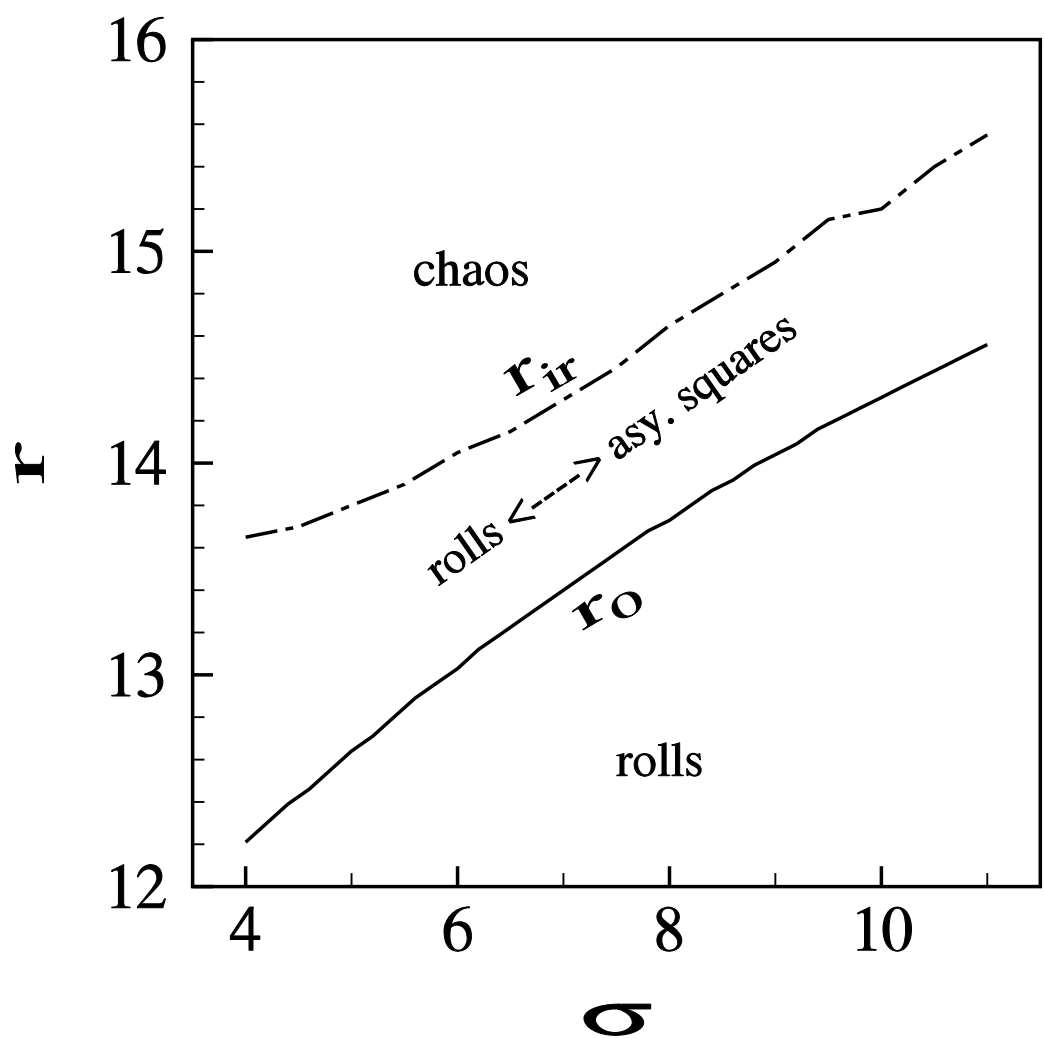
Figure 1. The region of $r$ and $\sigma$ space showing space rolls, asymmetric
squares and irregular solutions. The lower curve shows the values of reduced 
Rayleigh number $r_o$ above 2D rolls become unstable and time periodic 
asymmetric squares appear. The upper curve corresponds to the 
onset of irregular solutions in the model.

\noindent

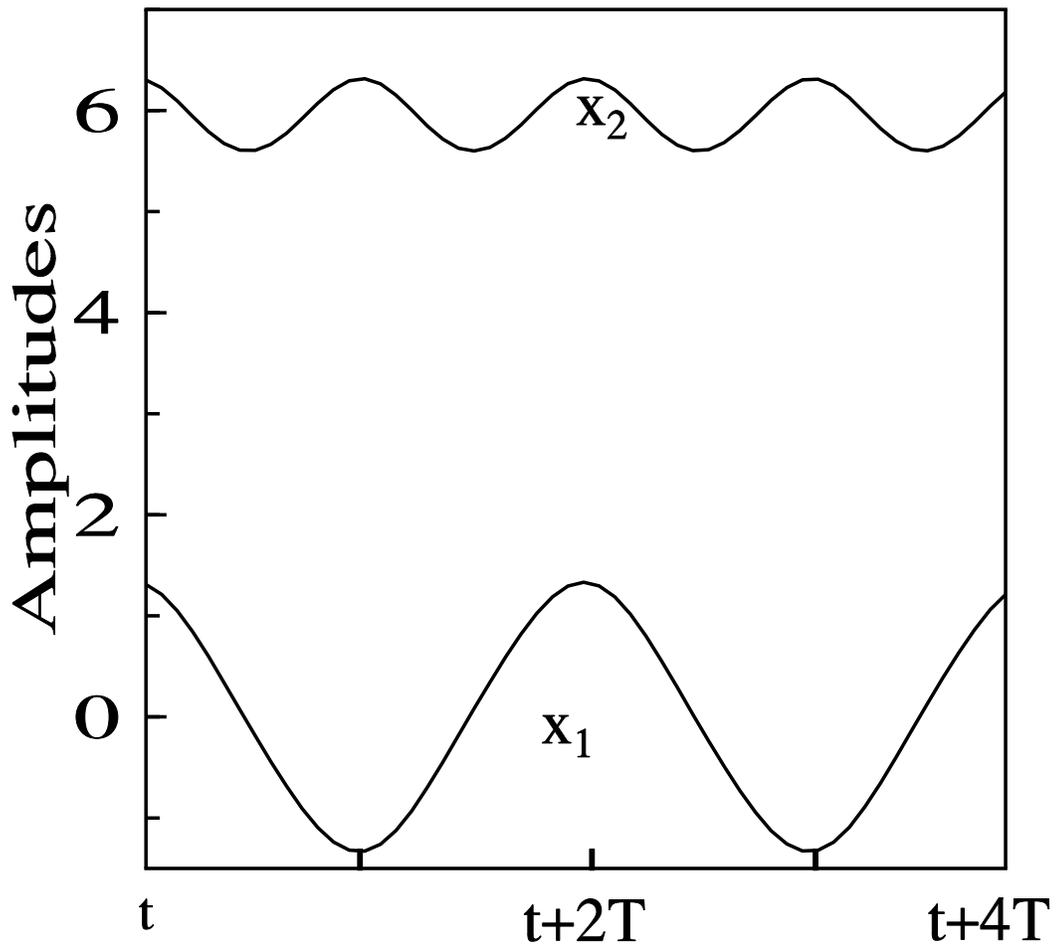
Figure 2. The  convective amplitudes $X_1$ and $X_2$ of new and old sets
of rolls respectively with respect to time for $r = 14.5$ and $r=14.7$
with value of $\sigma = 10$. The time period of the new set of rolls is
always double that for the old set of rolls, which became stable.

\noindent

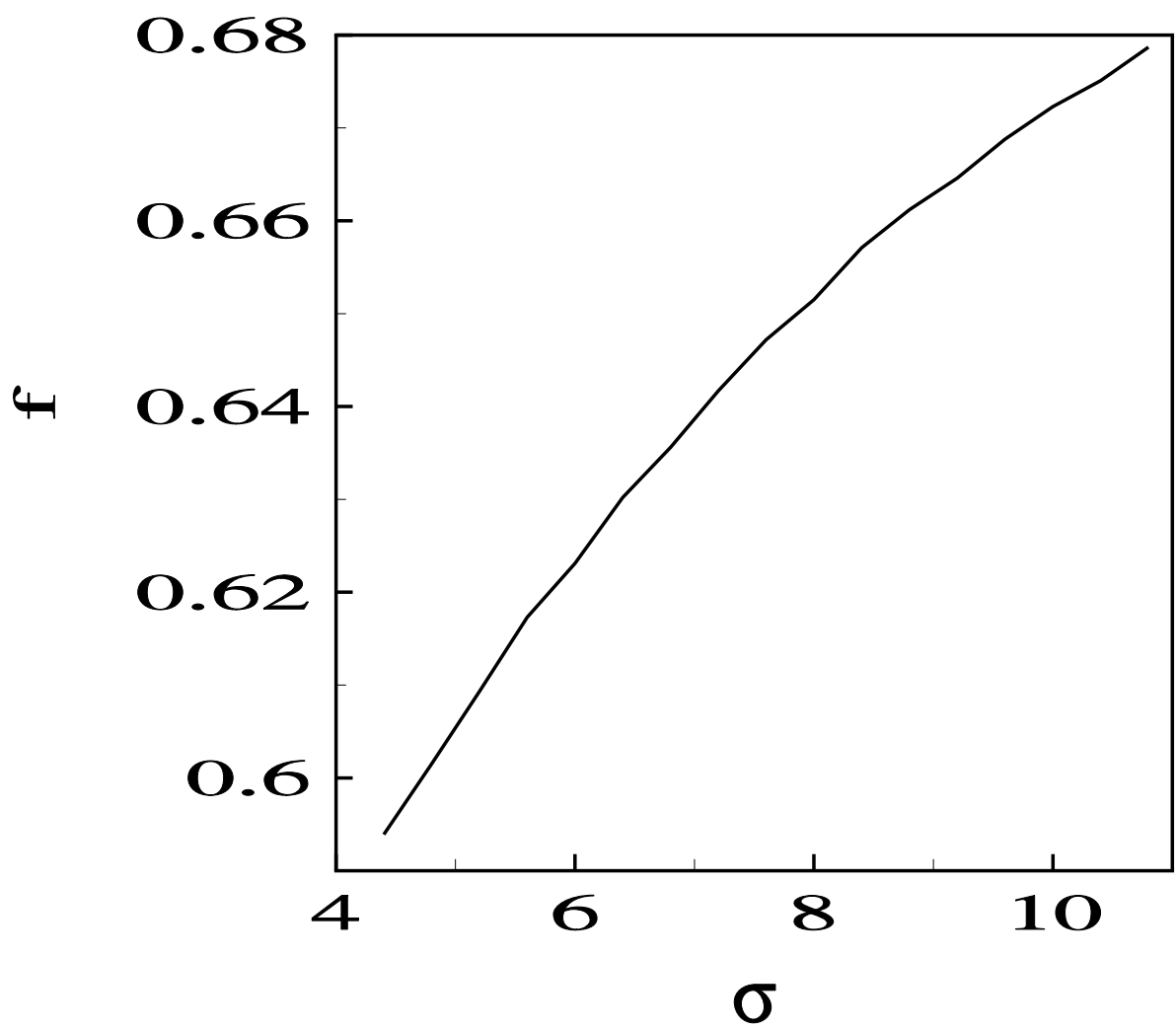
Figure 3. The plot of the frequency $f= 1/T$ of the amplitude $X_2$ as a 
function of Prandtl number $\sigma$ at the onset of secondary instability.

\noindent

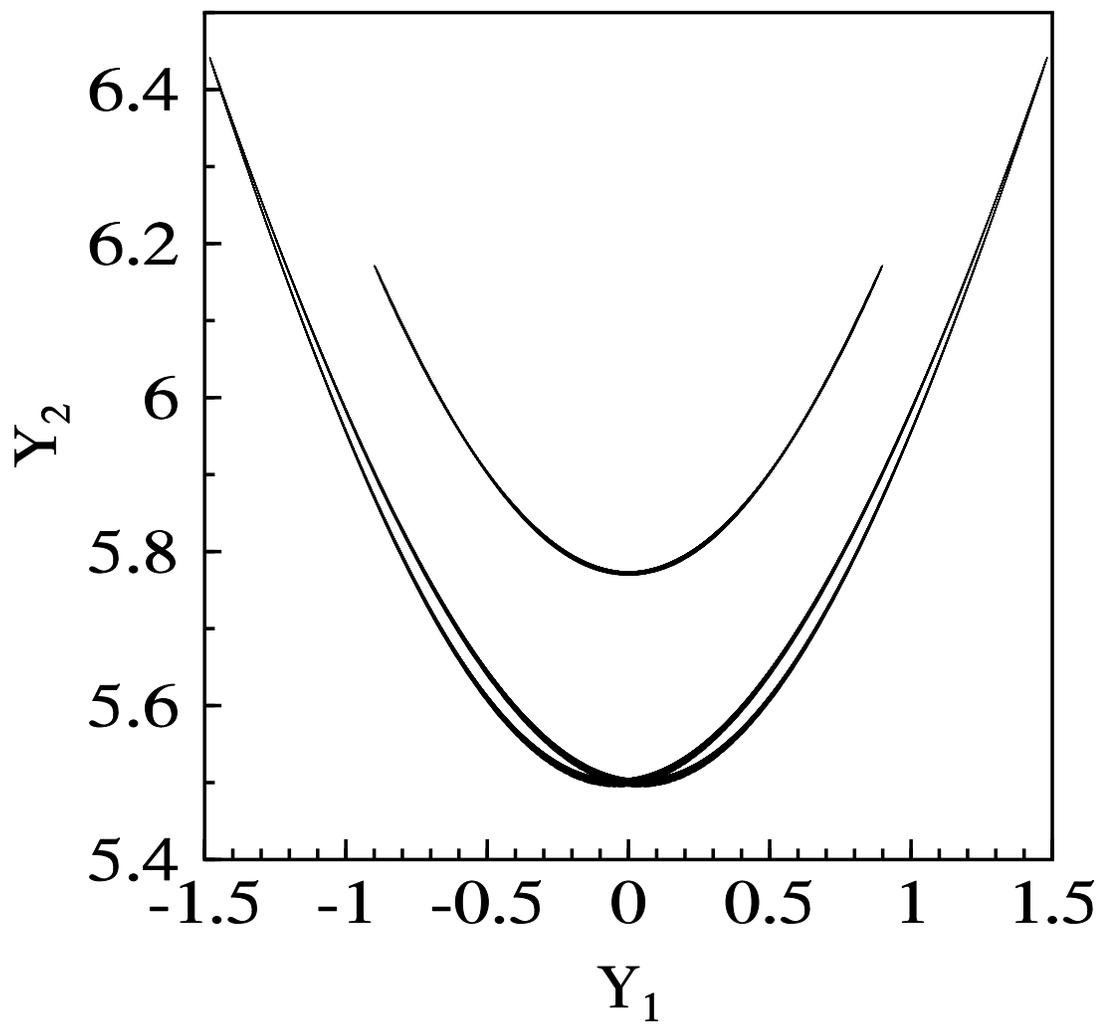
Figure 4. The projection of limit cycle on $Y_1 - Y_2 $ plane for
different values of $r$ for $\sigma =10$. Upper and lower plots correspond
to $r=14.5$ and $r=14.7$ respectively. The mean of 2D roll mode ($Y_2$) 
decreases and the perturbative amplitude $Y_1$ increases with increasing $r$. 

\noindent

Figure 5. Time-periodic sequence of asymmetric squares and rolls 
in shadowgraph for $r=14.7$ and $\sigma=10$. Shadowgraphs a,b,c correspond
respectively to  maximum, zero and minimum values of $Y_1$ in Fig. 4.   

\newpage

\begin{center}
{\bf REFERENCES}
\end{center}
\begin{enumerate}

\item{Schl\"uter, A., Lortz, D. and Busse, F., {\it J. Fluid Mech.} {\bf 23},
 129 (1965).}

\item{Thual, O., {\it J. Fluid Mech.} {\bf 240}, 229 (1992); see also
 Kumar, K., Fauve, S. and Thual, O., {\it J. Phys. II France} {\bf 6},
 945 (1996).}

\item{Whitehead, J. A. and Parsons, B., {\it Geophys. Astrophys. Fluid Dyn.}
 {\bf 9}, 201 (1978).}

\item{Assenheimer, M. and Steinberg, V., {\it Phys. Rev. Lett.} {\bf 76},
756 (1996).}

\item{Clever, R. M. and Busse, F. H., {\it Phys. Rev. E} {\bf 53},
R2037 (1996).}

\item{Busse, F. H. and Clever, R. M., {\it Phys. Rev. Lett.} {\bf13}, 341(1998).}

\item{Lorenz, E. N., {\it J. Atmos. Sci.} {\bf 20}, 130 (1963).}

\item{Moses, E. and Steinberg, V. {\it Phys. Rev. Lett.} {\bf 57}, 2018
(1986).}

\item{M\"uller, H. W. and L\"ucke, M. {\it Phys. Rev. A} {\bf 38}, 2965
(1988).}

\end{enumerate}

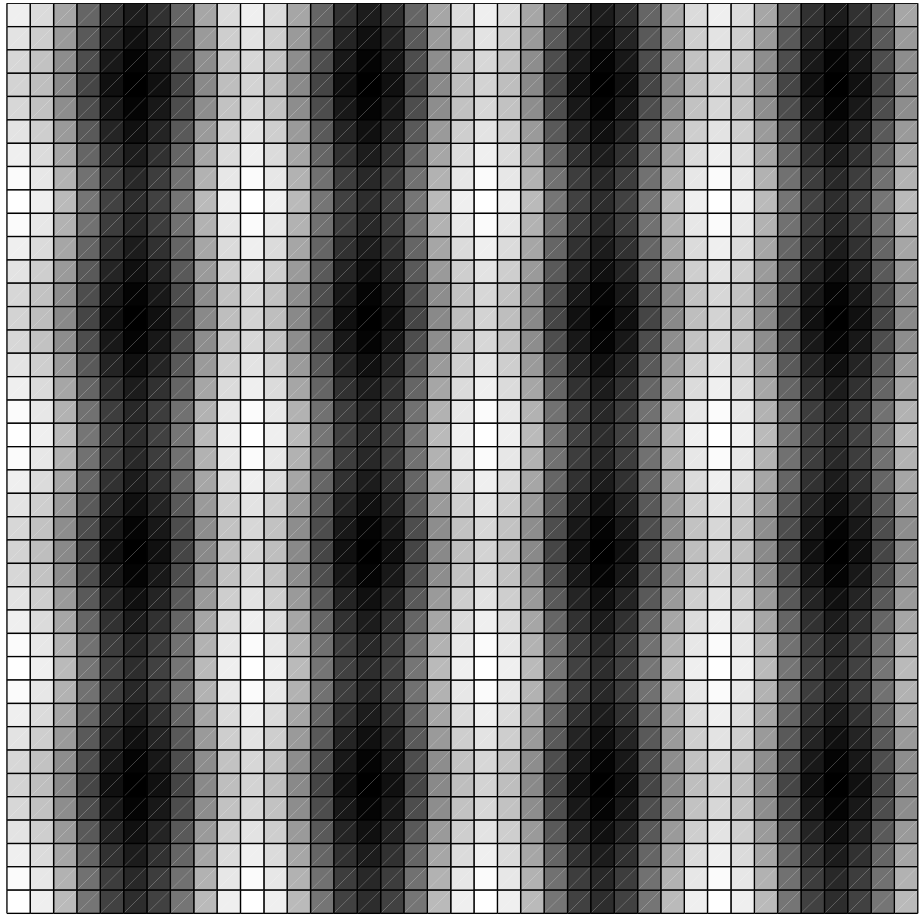
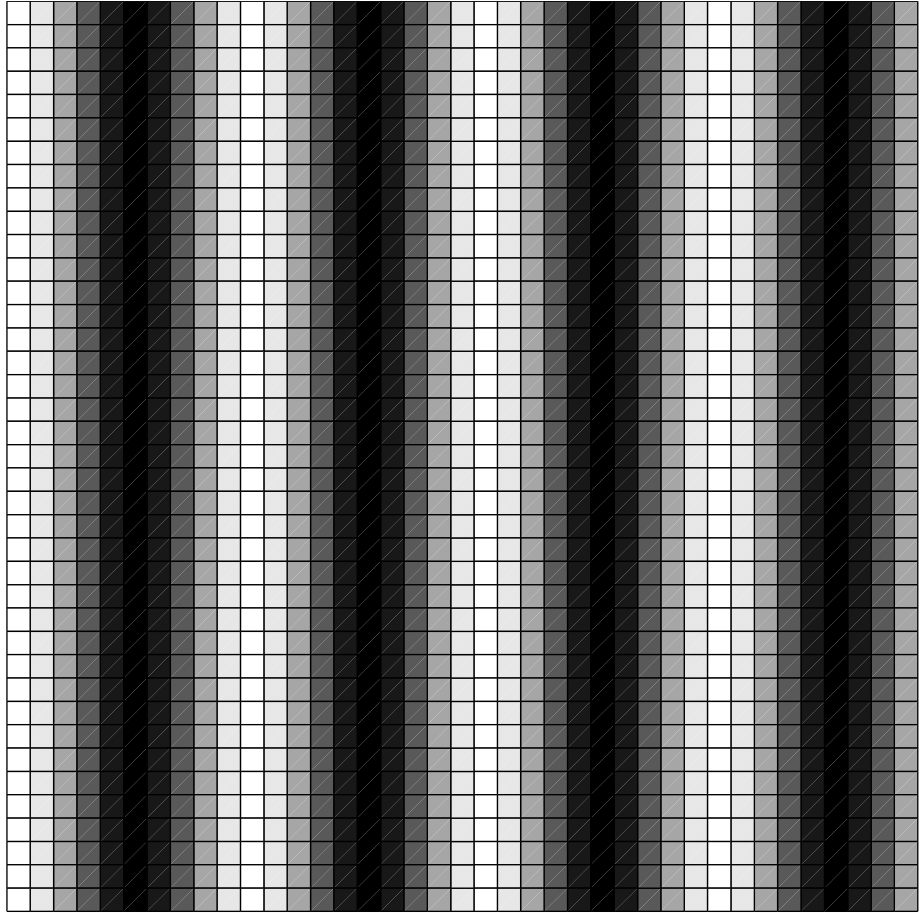
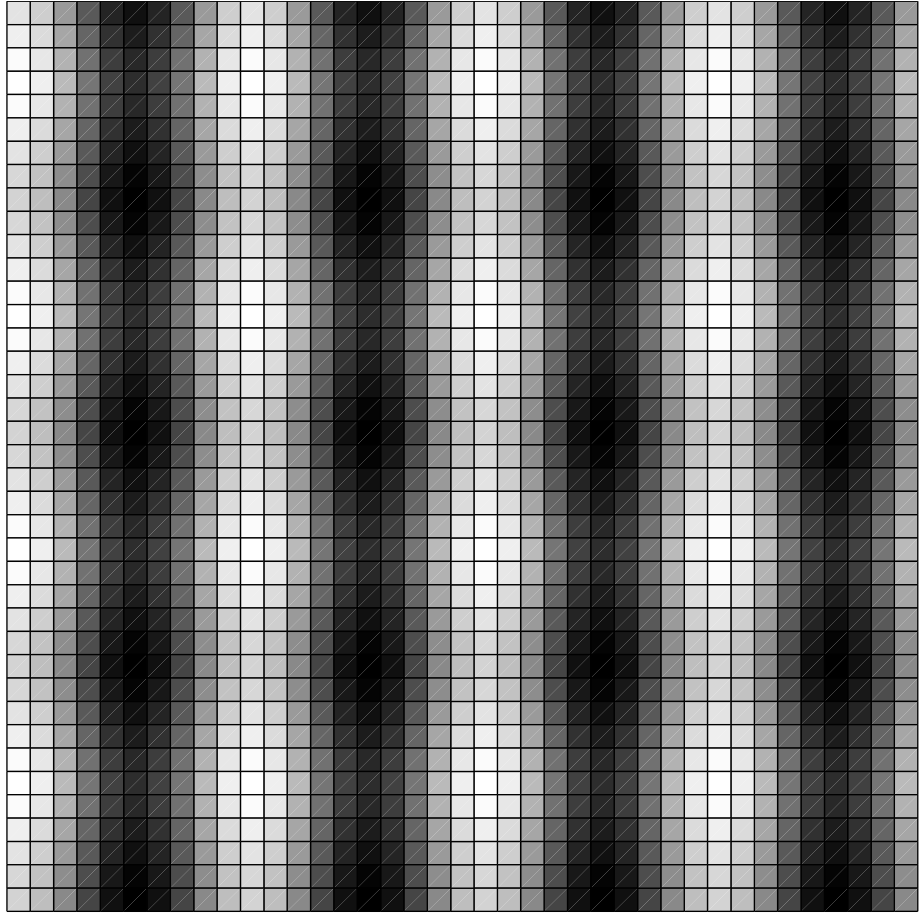
\end{document}